\documentclass[12pt]{iopart}
\usepackage{graphicx}

\begin{document}

\title[The dynamics on scales below the localization length]%
{Transport in the $3$-dimensional Anderson model: \\ an analysis of the dynamics on scales below the localization length}%

\author{Robin Steinigeweg$^1$, Hendrik Niemeyer$^2$, and Jochen Gemmer$^2$}%

\address{$^1$ Institut f\"ur Theoretische Physik, Technische Universit\"at Braunschweig,\\ Mendelsohnstrasse 3, D-38106 Braunschweig, Germany}%
\address{$^2$ Fachbereich Physik, Universit\"at Osnabr\"uck,\\ Barbarastrasse 7, D-49069 Osnabr\"uck, Germany}%

\ead{\mailto{r.steinigeweg@tu-bs.de}, \mailto{jgemmer@uos.de}}%

\begin{abstract}
Single-particle transport in disordered potentials is investigated
on scales \emph{below} the localization length. The dynamics on those
scales is concretely analyzed for the $3$-dimensional Anderson
model with Gaussian on-site disorder. This analysis particularly
includes the dependence of characteristic transport quantities on
the amount of disorder and the energy interval, e.g., the mean
free path which separates ballistic and diffusive transport
regimes. For these regimes mean velocities, respectively diffusion
constants are quantitatively given. By the use of the Boltzmann
equation in the limit of weak disorder we reveal the known
energy-dependencies of transport quantities. By an application of
the time-convolutionless (TCL) projection operator technique in
the limit of strong disorder we find evidence for much less
pronounced energy dependencies. All our results are partially
confirmed by the numerically exact solution of the time-dependent
Schr\"odinger equation or by approximative numerical integrators.
A comparison with other findings in the literature is additionally
provided.
\end{abstract}





\maketitle

\tableofcontents

%
%

\section{Introduction} \label{introduction}
Any solid contains disorder: Either there are impurities, vacancies, and
dislocations in an otherwise ideal crystal lattice. Or there is no lattice
structure at all. An abstract quantum system which is commonly used as a
paradigm for transport in real disordered solids is the Anderson model
\cite{anderson1958}. In its probably simplest form without particle-particle
interactions (electron-electron, electron-phonon, etc.) the Hamiltonian may be
written as
\begin{equation}
\hat{H} = \sum_{\vec{r}} \epsilon^{}_{\vec{r}} \,\,
\hat{a}^\dagger_{\vec{r}} \, \hat{a}^{}_{\vec{r}} +
\sum_{\textnormal{NN}} \hat{a}^\dagger_{\vec{r}_1}
\hat{a}^{}_{\vec{r}_2} \, , \label{H}
\end{equation}
where $\hat{a}_{\vec{r}}$ and $\hat{a}^\dagger_{\vec{r}}$ denote the
usual annihilation, respectively creation operators; $\vec{r}$ labels the sites
of a $d$-dimensional (cubic) lattice; NN indicates a sum over nearest neighbors
$\vec{r}_1$ and $\vec{r}_2$; and $\epsilon_{\vec{r}}$ represent independent
random numbers, e.g., according to a Gaussian distribution with mean $\langle
\epsilon_{\vec{r}} \rangle = 0$ and variance $\langle \epsilon_{\vec{r}_1} \,
\epsilon_{\vec{r}_2} \rangle = \delta_{\vec{r}_1, \vec{r}_2} \, \sigma^2$. Even
though such a distribution is considered throughout this work, the random
numbers can be realized according to a Lorentzian, box, or binary distribution
as well \cite{kramer1993, grussbach1995, slevin1999}. In all cases disorder is
implemented in terms of a random on-site potential. (Random hopping coefficients
are sometimes taken into account, too.)
\\
In the presence of such a disorder, $\sigma \neq 0$, the eigenstates of the
Hamiltonian are no longer given by Bloch functions: Instead the eigenstates are
not necessarily extended over the whole lattice and can become localized in
configuration space, i.e., the envelope of a wavefunction decays exponentially
on a finite localization length \cite{kramer1993, lee1985}. The finiteness of
the localization length is one manifestation and, say, definition of the
localization phenomenom. (There certainly are other mathematical definitions of
localization, e.g., the finiteness of the inverse participation number, the
independence of eigenvalues from boundary conditions, etc.~\cite{kramer1993})
This phenomenon and particularly its impact on transport have intensively been
studied for the Anderson model \cite{anderson1958, kramer1993, grussbach1995,
slevin1999, lee1985, abouchacra1973, abrahams1979, markos2006, brndiar2008,
lherbier2008}.
\\
For the lower dimensional cases, $d = 1$ and $d = 2$, all eigenstates of the
Hamiltonian feature finite localization lengths for arbitrary (non-zero) values
of $\sigma$ (except for situations with short-range correlated disorder, e.g.,
as realized in the random dimer model \cite{dunlap1990, bellani1999}). Therefore
in the thermodynamic limit, i.e., with respect to the infinite length scale an
insulator is to be expected. Of particular interest is the $3$-dimensional case,
as considered in the work at hand. Here, a mobility edge, i.e., a certain cross-over
energy separates the spatially localized from the spatially extended wavefunctions
in energy space \cite{kramer1993, grussbach1995, lee1985}. When the amount of disorder
is increased, the mobility edge goes above the Fermi level and a metal-to-insulator
transition is induced at zero temperature, still with respect to the infinite length
scale. When $\sigma$ is further increased above the critical disorder ($\sigma_C
\approx 6$ for a Gaussian distribution \cite{kramer1993, grussbach1995, slevin1999}),
all eigenstates become localized and an insulator is to be expected for each
temperature (without particle-particle interactions). The Anderson
metal-to-insulator transition is widely believed to be continuous without a
minimum conductivity, e.g., as supported by the one-parameter scaling theory of
localization \cite{kramer1993, lee1985, abrahams1979}.
\\
Our work, other than most of the pertinent literature, focuses on the dynamics
on scales \emph{below} the localization length. We particularly intend to
analyze the dynamics on those scales comprehensively as a function of energy and
disorder. Qualitatively, our analysis allows to identify two regimes of length
scales which are purely ballistic and strictly diffusive (rather than
superdiffusive, subdiffusive, or anything else). It generally is a challenge to
theoretically confirm reliably the ``presence of diffusion'' in strongly
disordered and/or interacting quantum systems. Quantitatively, our analysis
enables the evaluation of mean velocities, respectively diffusion coefficients
for a wide range of disorders between zero and the vicinity of $\sigma_C$.
Such a detailed knowledge about diffusion constants appears to be important,
especially since dc-conductivities are directly related by the Einstein
relation, at least for $\sigma < \sigma_C$. In the limit of strong disorder
diffusion coefficients have been suggested in the literature by the numerical
study of Green's functions for very few disorders and a single energy at the
spectral middle solely \cite{markos2006, brndiar2008}. The dependence of diffusion
constants on energy is usually discussed in the limit of weak disorder only.
However, also in that limit, we demonstrate that energy dependencies are much
richer than common approximations for a free electron gas \cite{kramer1993}.
\\
The work at hand is structured as follows: First of all we provide a
qualitative picture of the dynamics on scales below the localization
length in \sref{qualitative}. Then this qualitative picture is
subsequently developed and quantitatively confirmed in the whole
\sref{quantitative}: The limit of weak disorder is firstly analyzed
in \sref{weak} by the use of the Boltzmann equation \cite{peierls1965,
kadanoff1962, cercignani1988, bartsch2010}. The limit of
strong disorder is afterwards investigated by an application of a
method which is mainly based on the time-convolutionless (TCL)
projection operator technique \cite{chaturvedi1979, breuer2007, steinigeweg2007,
michel2007, steinigeweg2009}. In \sref{strong} the method as such
is introduced and its predictions on the dynamics are presented. The
validity range of these predictions is analytically discussed in
\sref{validity} and numerically verified in \sref{numerical}. We
finally close with a summary and conclusion in \sref{summary}.

%
%

\section{Qualitative picture of the dynamics} \label{qualitative}

\begin{figure}[htb]
\centering
\includegraphics[width=0.6\linewidth]{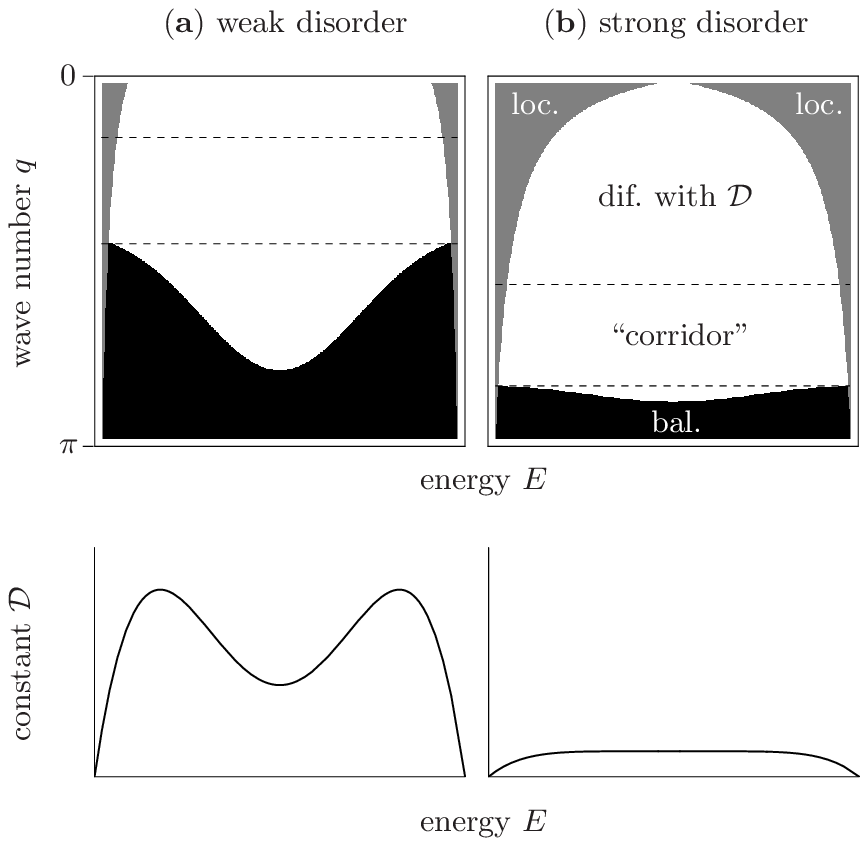}
\caption{Sketch for the dependence of transport on the wave number
$q$ and the energy $E$ for ({\bf a}) weak disorder and ({\bf b})
strong disorder. Both top panels indicate the rough position of
localized (loc., gray), diffusive (dif., white) and ballistic (bal.,
black) regimes. Particularly, a possible ``corridor'' of wave
numbers is indicated (dashed lines), where transport is diffusive
for almost all energies. Both bottom panels display suggestions for
qualitatively different energy dependencies of the diffusion
constant.} \label{figure_sketch}
\end{figure}

In the present section we intend to provide firstly a qualitative
picture of the dynamics on scales below the localization length.
This qualitative picture essentially summarizes the findings of the
methods which are introduced in detail and applied concretely in the
following sections. In particular we emphasize the main conclusions
of the work at hand and discuss these conclusions in the context of
known results in the literature. In this way we also give a
comprehensive summary for the readers which are not primarily
interested in the methodic details. Apart from that the summary
certainly makes the line of thoughts in the subsequent sections more
plainly.
\\
The above mentioned qualitative picture of the dynamics is
illustrated in \fref{figure_sketch}.
In this figure a sketch for the dependence of transport  on the
length scale $l$, respectively wave number $q$ and the energy $E$ is
shown for the two cases of ({\bf a}) weak disorder and ({\bf b})
strong disorder. For both cases the sketch indicates the rough
position of the different transport regimes, namely, localized
(loc., gray), diffusive (dif., white), and ballistic (bal., black).
\\
It is well known that in the limit of weak disorder the localization
phenomenon is restricted to the borders of the spectrum solely. Deep
in the outer tails of the spectrum the states are localized on a
single lattice site, whereas the overwhelming majority of all
states, i.e., not only the ones from the spectral middle, is still
extended. Thus, as displayed in \fref{figure_sketch}, the
localization length (envelope of gray areas) is not a closed curve
in the $(q,E)$-space. The energies which separate localized and
non-localized regimes at $q = 0$ (points between gray and white
areas at $q = 0$) are the mobility edges. And in fact, much work has
been devoted to the concrete position of the mobility edges \cite{kramer1993,
grussbach1995}. Only for energies between the mobility edges there is a
conductor at the infinite length scale, otherwise there is an insulator at
that length scale, of course. But, as already mentioned before,
insulating behavior is practically absent in the limit of weak
disorder, e.g., the Fermi level is much larger than the lower mobility
edge.
\\
While insulating behavior appears at rather large length scales
above the localization length, ballistic behavior occurs at
comparatively small length scales below the mean free path (envelope
of black area). Here, (quasi-)particles are not scattered and move
freely with mean (group) velocities, e.g., as routinely evaluated in
the framework of standard solid state theory. The latter free motion
is reflected in the term ballistic and is typical for an ideal
conductor. The mean free path, as drawn in \fref{figure_sketch},
appears to be a contra-intuitive curve in the $(q,E)$-space, since
it is smaller for states from the spectral middle than for states
from the borders of the spectrum. However, we demonstrate in
\sref{weak} that in the limit of weak disorder such a curve results
from the Boltzmann equation.
\\
Apparently, for weak disorder there is the practically unbounded
regime above the mean free path where transport is neither
insulating nor ballistic. This is the regime where transport is
generally expected to be diffusive, i.e., at that length scales one
expects a normal conductor. Particularly, \fref{figure_sketch} marks
a ``corridor'' of wave numbers (dashed lines) with diffusive
dynamics at almost all energies. (The notion of a diffusive corridor
becomes helpful for later argumentations in the context of strong
disorder where the existence of such a corridor is anything else
than obvious.) From a mere theoretical point of view it is a
challenge to concisely show that the dynamics is in full accord with
a diffusion equation. But in \sref{weak} we demonstrate by the use
of the Boltzmann equation that the dynamics is indeed diffusive and
further evaluate quantitatively the diffusion coefficient as a
function of energy. As indicated in \fref{figure_sketch}, its
dependence on energy seems to be as contra-intuitive as the one of
the mean free path and strongly differs in the details from the
approximations according to a free electron gas.
\\
For strong disorder the localization phenomenon is much more
pronounced. When the amount of disorder is increased, the localized
regimes gradually expand towards small length scales and towards
energies in the middle of the spectrum as well, see
\fref{figure_sketch}. On the one hand the already non-extended
states become localized on smaller and smaller length scales. On the
other hand more and more of the before extended states become
localized at all. Hence, the mobility edges move closer to each
other and eventually meet, once the critical disorder is reached.
Then all states are localized and transport at the infinite length
scale vanishes completely, i.e., at all energies. Therefore much
work has addressed the concrete evaluation of the critical disorder
\cite{kramer1993, grussbach1995, slevin1999}. However, even above the
critical disorder, transport takes place below the localization length,
of course.
\\
It is a priori not clear whether or not the dynamics below the
localization length is still in good agreement with a diffusion
equation in the limit of strong disorder, both below and above the
critical disorder. But by the use of a method which is based on the
TCL projection operator technique we demonstrate in \sref{strong}
that there also exists a corridor of wave numbers where the dynamics
is indeed diffusive at almost all energies, at least as long as the
amount of disorder does not become too strong. In particular the
diffusion constant within this corridor does not substantially
depend on energy, see \fref{figure_sketch}. In fact, only if the
dynamics for a certain wave number is not governed by a significant
energy dependence, the method makes a definite conclusion, otherwise
no information results except for the strong energy dependence of
the dynamics, e.g., the method can not distinguish between highly
energy dependent diffusion coefficients and non-negligible localized
contributions. However, once a diffusive corridor of wave numbers
with a single diffusion constant is reliably detected, it is natural
to assume that the diffusion coefficient does not change, when this
corridor is left. (Per definition diffusion coefficients should
not depend on the wave number). Or, in other words, we suggest
that in the limit of strong disorder the dynamics in the whole
diffusive regime is well described by an energy independent
diffusion constant. This diffusion constant is quantitatively
evaluated in \sref{strong} as a function of disorder.
\\
For the case of strong disorder the TCL-based method additionally
allows to characterize the ballistic regime, i.e., by the use of the
method the mean free path and the mean velocity can be also
evaluated. Similarly, these quantities are found to be approximately
independent from energy. This observation suggests that in the limit
of strong disorder the whole dynamics below the localization length
is not governed by significant energy dependencies. Of course, in a
sense this suggestion disagrees with the observations for the case
of weak disorder. Nevertheless, in the sections \ref{weak} and
\ref{strong} the disagreement is subsequently resolved, both
qualitatively and quantitatively.

%
%

\section{Quantitative results} \label{quantitative}

%
%

\subsection{Weak disorder: Boltzmann equation} \label{weak}

\begin{figure}[htb]
\centering
\includegraphics[width=0.6\linewidth]{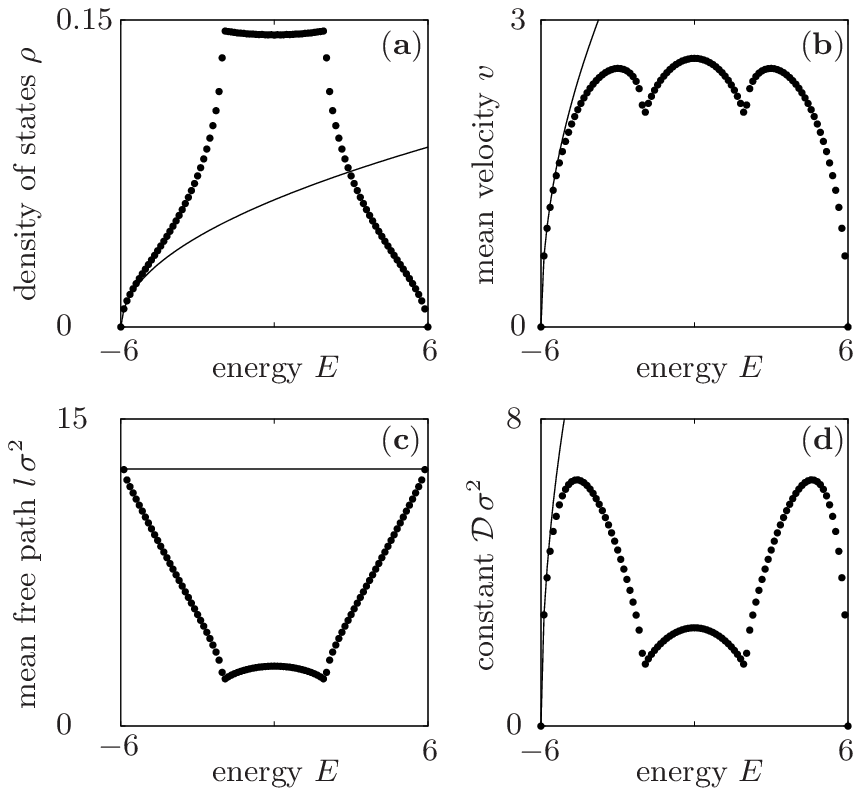}
\caption{Energy-dependencies in the limit of weak disorder:
({\bf a}) density of states, ({\bf b}) mean velocity, ({\bf c}) mean
free path as well as ({\bf d}) diffusion constant (according to the
Boltzmann equation). The numerical data (circles) is extracted from a
cube with $1000^3$ lattice sites. The known approximations for low
energies (solid lines) are indicated for comparison.}
\label{figure_boltzmann}
\end{figure}

\begin{figure}[htb]
\centering
\includegraphics[width=0.5\linewidth]{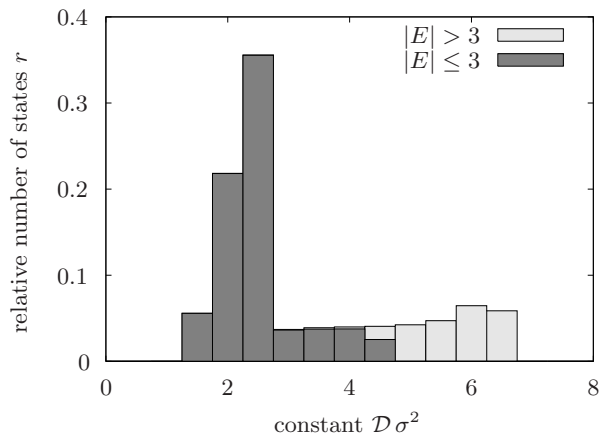}
\caption{The relative number of states $r$ corresponding to a
diffusion constant $\cal D$ (according to the Boltzmann equation),
cf.~Fig.~\ref{figure_boltzmann}. The bars of the histogram visualize
contributions from energies $|E| > 3$ (light-colored area) and from
energies $|E| \leq 3$ (dark-colored area).}
\label{figure_distribution}
\end{figure}

In the present section we are going to investigate the dynamics in the limit of
weak disorder. In that limit there certainly is a large variety of different
approaches which all treat the disorder as a small perturbation to the clean
Hamiltonian. Here, we briefly review on one class of these approaches, namely,
the mapping of the quantum dynamics onto Boltzmann equations \cite{peierls1965,
kadanoff1962, cercignani1988, bartsch2010}. Different approaches to such a map
rely on different assumptions and/or approximation schemes which are not entirely
free of their own subtleties. However, the particle velocities that eventually
enter the Boltzmann equation are routinely taken from the clean (unperturbed)
Hamiltonian. To this end the clean Hamiltonian has to be diagonalized at first.
Routinely, this diagonalization can be done by the application of the Fourier
transform. Then the Hamiltonian takes on the form
\begin{equation}
\hat{H} = \sum_{\vec{q}} \, E_{\vec{q}} \,\,
\hat{a}^{\dagger}_{\vec{q}} \, \hat{a}^{}_{\vec{q}} \, ,
\end{equation}
where $\hat{a}^{\dagger}_{\vec{q}}$, $\hat{a}^{}_{\vec{q}}$ are
creation, annihilation operators for (quasi-)particles with the wave
vector $\vec{q}$, i.e., $q_i = 2 \pi \, k_i / N$, $k_i = 0, \ldots,
N-1$. The corresponding dispersion relation reads
\begin{equation}
E_{\vec{q}} = \sum_{i = 1}^3 \, 2 \, (1 - \cos q_i) \approx | \,
\vec{q} \, |^2 \, .
\end{equation}
(Now and in the following the indicated approximations hold true for
sufficiently small $| \, \vec{q} \, |$, respectively low energies and are well
known from the free electron gas.) As long as disorder is absent, the
(quasi-)particles are not scattered and may be said to move freely with the
(group) velocities which are determined by the derivative of the dispersion
relation, namely,
\begin{equation}
v_{\vec{q}} = | \, \nabla_{\! \vec{q}} \, E_{\vec{q}} \, | = 2 \,
\sqrt{\sum_{i = 1}^3 \, \sin^2 q_i} \approx 2 \, | \, \vec{q} \, |
\, .
\end{equation}
Whithin such a Boltzmann equation framework disorder takes the role of a set of
impurities from which the (quasi-)particles are scattered after, say, a mean
free time $\tau(E)$, respectively mean free path $l(E)$. Therefore disorder
essentially gives raise to a linear collision term, i.e., a rate matrix which
describes the transitions between different (quasi-)momentum eigenstates.
Generally, diffusion coefficients may be computed based on the inverse of this
rate matrix. However, since the disorder of the Anderson model (statistically)
features full spherical symmetry, a relaxation time approximation turns out to
be exact, even though the dispersion relation does not feature full spherical
symmetry \cite{bartsch2010}. Following this approach, the diffusion coefficient
may be cast into the basic form
\begin{equation}
{\cal D}(E) = \frac{1}{3} \, v(E) \, l(E) \, , \quad l(E) = v(E) \,
\tau(E) \, , \quad \tau(E) = \frac{1}{2 \pi \, \rho(E) \, \sigma^2} \, ,
\label{D_boltzmann}
\end{equation}
where $v(E)$ denotes a mean velocity which is obtained from an average over all
$\vec{q}$ featuring a certain  energy $E$, i.e.,
\begin{equation}
v(E) = \langle \, v_{\vec{q}} \, \rangle_{\displaystyle \{ \vec{q}
\, | \, E_{\vec{q}} = E \}} \approx 2 \, \sqrt{E} \, .
\label{v_boltzmann}
\end{equation}
Furthermore, $\rho(E)$ expresses the density of states normalized to the
volume, i.e., 
\begin{equation}
\rho(E) = \frac{1}{N^3} \,
\frac{\textnormal{d}Z(E)}{\textnormal{d}E} \approx \frac{\sqrt{E}}{4
\pi^2} \label{rho_boltzmann}
\end{equation}
w.r.t.~the clean Hamiltonian. As a first observation, diffusion coefficients and
mean free paths are inversely proportional to the amount of disorder, at least
within the Boltzmann equation approach at hand. Due to the above mentioned subtelties
of the mapping itself it is hard to give a detailed estimate for the regime of its
applicability. However, disorder should generally be substantially smaller than
regular hopping, i.e, $\sigma \ll 1$.
\\
Inserting the approximations for low energies in \eref{v_boltzmann},
\eref{rho_boltzmann} into \eref{D_boltzmann} yields
\begin{equation}
{\cal D}(E) \approx \frac{8 \pi \, \sqrt{E}}{3 \, \sigma^2} \, ,
\quad l(E) \approx \frac{4 \pi}{\sigma^2} \, .
\end{equation}
This result coincides with the one in \cite{kramer1993} which is
found therein by the use of Green's functions. However, in order to
obtain the full energy dependencies of ${\cal D}(E)$ and $l(E)$ we
numerically evaluate \eref{v_boltzmann} and \eref{rho_boltzmann} in
\fref{figure_boltzmann} ({\bf a}) and ({\bf b}). Note that the
evaluation can be done for very large lattices, e.g., $N = 1000$,
since exact diagonalization is not involved.
Obviously, the approximations for \eref{v_boltzmann} and
\eref{rho_boltzmann} are valid for very low energies solely, i.e.,
in the outer tails of the density of states. The actual curves
differ strongly in the details. As a consequence the curves for
${\cal D}(E)$ and $l(E)$, as displayed in \fref{figure_boltzmann}
({\bf c}) and ({\bf d}), show interesting features, too.
Particularly, the maximum diffusion coefficient is not located at
the middle of the spectrum ($E = 0$). Instead two distinct maxima
are observed at positions which are closer to the borders of the
spectrum ($E \approx 4.5$).
\\
The curve for ${\cal D}(E)$ seems to already indicate that the
overall dynamics of all energy regimes, i.e., the dynamics at high
temperatures can not be described as diffusive with a single
diffusion coefficient. However, for a definite conclusion the ${\cal
D}(E)$-curve has to be weighted with the density of states
$\rho(E)$, of course. Therefore in \fref{figure_distribution} the
relative number of states $r$ is shown which contribute to a certain
diffusion constant.
In a sense $r$ is again a density of states but now in the space of
diffusion coefficients. Apparently, the majority of all states
corresponds to diffusion constants which are rather close to ${\cal
D} \approx 2.5 / \sigma^2$, i.e., the value for $E = 0$. These
states are also located around the middle of the spectrum, as
indicated in \fref{figure_distribution}. But there is a relevant
number of states from the outer parts of the spectrum which
contribute to larger values of $\cal D$. Remarkably, in these parts
the number of states with smaller values of $\cal D$ is negligible.
However, \fref{figure_distribution} clearly demonstrates that the
overall dynamics of all energy regimes is not diffusive with a
single diffusion coefficient.
\\
The situation may change, when the limit of weak disorder is
slightly left, i.e., when the above predictions of the Boltzmann
equation begin to break down. Obviously, the breakdown begins for
the states from the borders of the spectrum, since these states are
the first which become eventually localized. (The Boltzmann equation
does simply not predict localization).  At this point the
predictions for the states from the spectral middle are still
unaffected, of course. On that account it may happen that the large
values of $\cal D$ in \fref{figure_distribution} are gradually moved
towards ${\cal D} \approx 2.5 / \sigma^2$ such that $r$ finally
becomes more or less peaked at this position. In that case the
dynamics is governed by a single diffusion constant. So far, this
line of thoughts is a mere assumption. Even if the assumption was
correct, it would be entirely unclear whether or not this assumption
has some impact on a situation with strong disorder, i.e., beyond
any validity of the Boltzmann equation.
\\
In the following sections \ref{strong} and \ref{validity} we
subsequently show for strong disorder that it appears to be indeed
justified to describe the dynamics below the localization length as
diffusive with a single diffusion coefficient. Surprisingly, this
diffusion constant is rather close to the value $2.5 / \sigma^2$,
wide outside the strict validity of the Boltzmann equation.

%
%

\subsection{Strong disorder: TCL projection operator technique} \label{strong}

\begin{figure}[htb]
\centering
\includegraphics[width=0.5\linewidth]{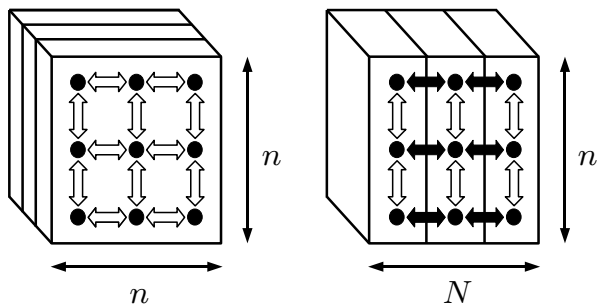}
\caption{A $3$-dimensional lattice which consists of $N$ layers with
$n \times n$ sites each. Only next-neighbor hoppings are taken into
account. Constants for intra-layer hoppings are set to $1$ (white
arrows), inter-layer hoppings are specified by another constant
$\lambda$ (black arrows).} \label{figure_model}
\end{figure}

\begin{figure}[htb]
\centering
\includegraphics[width=0.6\linewidth]{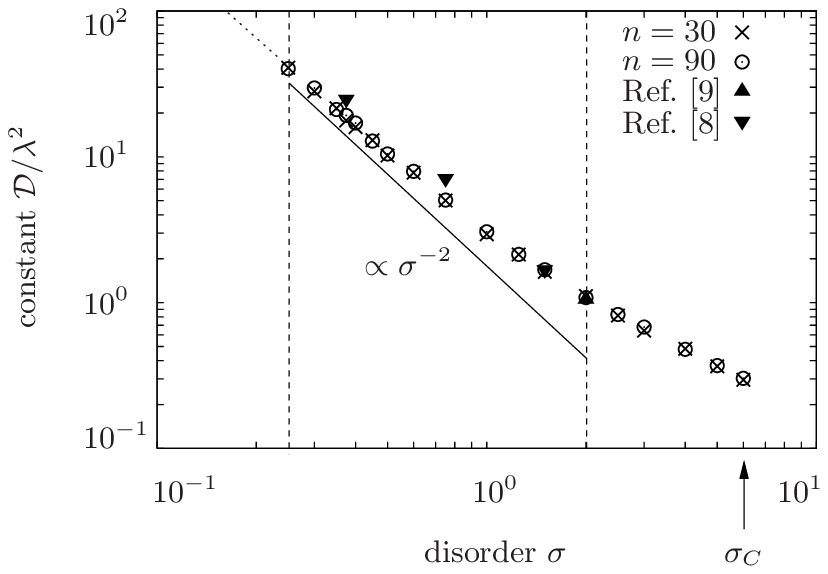}
\caption{Theoretical prediction of lowest order TCL for an
energy-independent diffusion constant $\cal D$ as a function of the
disorder $\sigma$ (Gaussian distribution) for the layer sizes $n =
30$ (crosses) and  $n = 90$ (circles). The theoretical prediction is
shown up to the critical disorder $\sigma_C \approx 6$ and the
validity range of this prediction is displayed (dashed lines). As a
guide for the eyes a proportionality to $\sigma^{-2}$ is indicated
(solid line). For comparison the value ${\cal D} = 1.05 \pm 0.10$
from \cite{brndiar2008} is shown (triangle), as found therein for $E
= 0$ and $\sigma = 2$ (Gaussian distribution). Further values for
$\cal D$ from \cite{markos2006} are displayed (triangles), as
obtained therein for $E = 0$ but for $W = 1$, $2$ and $4$ (box
distribution). In that case data for Gaussian and box distributions
are supposed to be simply convertible by $W/\sigma \approx 2.6$
(according to the ratio of the critical values). The prediction for
$E = 0$ according to the Boltzmann equation is also indicated
(dotted line).} \label{figure_D}
\end{figure}

Our approach in the limit of strong disorder is based on the
time-convolutionless (TCL) projection operator technique
\cite{chaturvedi1979,breuer2007} which has
already been applied to the transport properties of similar models
but without disorder, see \cite{steinigeweg2007,michel2007,steinigeweg2009}.
In its standard form this approach is restricted to the infinite temperature
limit. This limitation implies that energy dependencies are not resolved,
i.e., our results are to be interpreted as results on an overall behavior
of all energy regimes.
\\
As illustrated in \fref{figure_model},
we consider a $3$-dimensional lattice consisting of $N$ layers with
$n \times n$ sites each. The Hamiltonian of our model is almost
identical to \eref{H} with one single exception: All hopping terms
which correspond to hoppings between layers (black arrows in
\fref{figure_model}) are multiplied by some constant $\lambda$. This
multiplication is basically done due to technical reasons, see
below. However, for $\lambda = 1$ the Hamiltonian reduces to the
usual Anderson Hamiltonian \eref{H}.
\\
We now establish a coarse-grained description in terms of subunits (similarly
to \cite{weaver2006}):
At first we take all those terms of the Hamiltonian which only
contain the sites of the $\mu$th layer in order to form the local
Hamiltonian $\hat{h}_\mu$ of the subunit $\mu$. Thereafter all those
terms which contain the sites of adjacent layers $\mu$ and $\mu+1$
are taken in order to form the interaction $\lambda \, \hat{v}_\mu$
between neighboring subunits $\mu$ and $\mu+1$. Then the total
Hamiltonian may be also written as
\begin{equation}
\hat{H} = \hat{H}_0 + \lambda \, \hat{V} \, , \quad \hat{H}_0 =
\sum_{\mu = 0}^{N-1} \hat{h}_\mu \, , \quad \hat{V} = \sum_{\mu =
0}^{N-1} \hat{v}_\mu \, ,
\end{equation}
where we use periodic boundary conditions, e.g., we identify $\mu =
N$ with $\mu = 0$. The introduction of the additional parameter
$\lambda$ allows for the independent adjustment of the interaction
strength in this coarse-grained description. Since we are going to
work in the Dirac picture, the indispensable eigenbasis of
$\hat{H}_0$ may be found from the diagonalization of disconnected
layers.
\\
By $\hat{p}_\mu$ we denote the particle number operator of the
$\mu$th subunit, i.e., the sum of $\hat{a}^\dagger_{\vec{r}} \,
\hat{a}^{}_{\vec{r}}$ over all ${\vec{r}}$ of the $\mu$th layer.
Because the overall number of particles is conserved, i.e., $[ \,
\sum_\mu \hat{p}_\mu, \hat{H} \, ] = 0$ and no particle-particle
interactions are taken into account, we still restrict the
investigation to the one-particle subspace. The actual state of the
system is naturally represented by a time-dependent density matrix
$\rho(t)$, i.e., the quantity $p_\mu(t) = \textnormal{Tr} \{ \,
\rho(t) \, \hat{p}_\mu \, \}$ is the probability for locating the
particle somewhere within the $\mu$th subunit. The consideration of
these coarse-grained probabilities corresponds to the analysis of
transport along the direction which is perpendicular to the layers.
Instead of simply characterizing whether or not there is transport
at all, we analyze the full dynamics of the $p_\mu(t)$.
\\
The dynamical behavior of the $p_\mu(t)$ may be called diffusive, if
the $p_\mu(t)$ fulfill a discrete diffusion equation
\begin{equation}
\dot{p}_\mu(t) = {\cal D} \, [ \, p_{\mu-1}(t)- 2 \, p_\mu(t) +
p_{\mu+1}(t) \, ] \label{diffusion1}
\end{equation}
with some $\mu$- and $t$-independent diffusion constant $\cal D$. It
is a straightforward manner to show (multiplying \eref{diffusion1}
by $\mu$, respectively $\mu^2$, performing a sum over $\mu$ and
manipulating indices on the r.h.s.) that the spatial variance
\begin{equation}
\textnormal{Var}(t) = \sum_{\mu = 0}^{N-1} \mu^2 \, p_\mu(t) - \left
[ \, \sum_{\mu = 0}^{N-1} \mu \, p_\mu(t) \, \right ]^2
\label{variance}
\end{equation}
increases linearly with $t$, i.e., $\textnormal{Var}(t) = 2 \, {\cal
D} \, t$. Contrary, ballistic behavior is characterized by
$\textnormal{Var}(t) \propto t^2$, while insulating behavior
corresponds to $\textnormal{Var}(t) = \textnormal{const.}$, of
course.
\\
According to Fourier's work, diffusions equations are routinely
decoupled with respect to, e.g., cosine-shaped spatial density
profiles
\begin{equation}
p_q(t) = C_q \sum_{\mu = 0}^{N-1} \cos(q \, \mu) \, p_\mu(t) \, ,
\quad q = \frac{2 \pi \, k}{N} \, , \quad k = 0, 1, \ldots,
\frac{N}{2}
\end{equation}
and a yet arbitrary normalization constant $C_q$. Consequently,
\eref{diffusion1} yields
\begin{equation}
\dot{p}_q(t) = -2 \, (1 - \cos q) \, {\cal D} \, p_q(t) \, .
\label{diffusion2}
\end{equation}
Therefore, if the quantum model indeed shows diffusive transport,
all modes $p_q(t)$ have to relax exponentially. If, however, the
modes $p_q(t)$ are found to relax exponentially only for some regime
of $q$, the model is said to behave diffusively on the corresponding
length scale $l = \pi / q$. One might think of a length scale which
is both large compared to the mean free path (below that ballistic
behavior occurs, $\sigma^2(t) \propto t^2$) and small compared the
localization length (beyond that insulating behavior appears,
$\sigma^2(t) = \textnormal{const.}$).
\\
For our purposes, i.e., for the application of the TCL projection
operator technique, it is convenient to express the modes $p_q(t)$
as the  expectation values of respective mode operators $\hat{p}_q$,
namely,
\begin{equation}
p_q(t) = \textnormal{Tr} \{ \, \rho(t) \, \hat{p}_q \, \} \, , \quad
\hat{p}_q = C_q \sum_{\mu = 0}^{N - 1} \cos (q \, \mu) \,
\hat{p}_\mu \, ,
\end{equation}
where the normalization constants $C_q$ are now chosen such that
$\textnormal{Tr} \{ \, \hat{p}_q^2 \, \} = 1$. With this
normalization
\begin{equation}
{\cal P} \, \rho(t) = \textnormal{Tr} \{ \, \rho(t) \, \hat{p}_q \,
\} \, \hat{p}_q = p_q(t) \, \hat{p}_q \,
\end{equation}
defines a suitable projection (super)operator, because ${\cal P}^2 =
{\cal P}$. For those initial states $\rho(0)$ which satisfy ${\cal
P} \, \rho(0) = \rho(0)$, i.e., for harmonic density profiles the
TCL projection operator technique eventually leads to a differential
equation of the form
\begin{equation}
\dot{p}_q(t) = R_q(t) \, p_q(t) \, , \quad R_q(t) = \lambda^2 \,
R_{2,q}(t) + \lambda^4 \, R_{4,q}(t) + \ldots \label{tcl}
\end{equation}
which is a formally exact description for the dynamics at high
temperatures, since $\rho(0)$ is not restricted to any energy
subspaces. Apparently, the dynamics of $p_q(t)$ is controlled by a
time-dependent decay rate $R_q(t)$. This decay rate is given in
terms of a systematic perturbation expansion in powers of the
inter-layer coupling. (Concretely, for this model all odd orders
vanish.) At first we concentrate on the truncation of \eref{tcl} to
lowest order, i.e., to second order. But the fourth order is
considered afterwards in order to estimate the validity of this
second order truncation.
\\
According to \cite{breuer2007}, the TCL formalism routinely yields
the second order prediction 
\begin{equation}
\dot{p}_q(t) = \lambda^2 \, R_{2,q}(t) \, p_q(t) \, , \quad
R_{2,q}(t) = \int_0^t \textnormal{d} \Delta \, f_q(\Delta)
\end{equation}
with the two-point correlation function
\begin{equation}
f_q(\Delta) = \textnormal{Tr} \Big \{ [ \, \hat{V}(t), \hat{p}_q \, ]
\, [ \, \hat{V}(t'), \hat{p}_q \, ] \Big \} \, , \quad \Delta = t - t'
\, , \label{correlation1}
\end{equation}
where the time dependencies of operators are to be understood with
respect to the Dirac picture. The $q$-dependence in
\eref{correlation1} is significantly simplified under the following
assumption: The autocorrelation functions $\textnormal{Tr} \{ \,
\hat{v}_\mu(t) \, \hat{v}_\mu(t') \, \}$ of the local interactions
$\hat{v}_\mu$ should depend only negligibly on the layer number
$\mu$ (at relevant time scales). In fact, numerics indicate that
this assumption is well fulfilled (for the values of $\sigma$ which
are discussed here), once the layer sizes exceed ca.~$30 \times 30$.
Therefore first investigations may be based on the consideration of
an arbitrarily chosen junction of two layers. The local interaction
between these representative layers may be called $\hat{v}_0$. The
use of the above assumption simplifies \eref{correlation1} to
\begin{equation}
f_q(\Delta) \approx -2 \, (1 - \cos q) \, f(\Delta) \, , \quad f(\Delta) =
\frac{1}{n^2} \, \textnormal{Tr} \{ \, \hat{v}_0(t) \, \hat{v}_0(t')
\, \} \, , \label{correlation2}
\end{equation}
where the $q$-dependence enters solely as an overall scaling factor
\cite{steinigeweg2009}. As a consequence the second order prediction
at high temperatures reads
\begin{equation}
\dot{p}_q(t) \approx -2 \, (1 - \cos q) \, \lambda^2 \, R_2(t) \,
p_q(t) \, , \quad R_2(t) = \int_0^t \textnormal{d} \Delta \, f(\Delta)
\, . \label{tcl2}
\end{equation}
This equation is already very similar to \eref{diffusion2} but still
contains a time-dependent diffusion coefficient ${\cal D}(t) =
\lambda^2 \, R_2(t)$. However, it numerically turns out that
$f(\Delta)$ behaves like a standard correlation function, i.e., it
decays completely within some time scale $\tau_C$. After this
correlation time $f(\Delta)$ approximately remains zero and $R_2(t)$
takes on a constant value $R_2$, the area under the initial peak of
$f(\Delta)$. Numerics indicates that neither $\tau_C$ nor $R_2$ depend
substantially on $n$ (at least for $n > 30$) such that both $\tau_C$
and $R_2$ are essentially functions of $\sigma$. Since the
correlation time $\tau_C$ apparently is independent from $q$ and
$\lambda$, it is always possible to realize a relaxation time
\begin{equation}
\tau_R = \frac{1}{2 \, (1- \cos q) \lambda^2 R_2}
\end{equation}
which is much larger than $\tau_C$, e.g., in an infinitely large
system there definitely is a small enough $q$. For $\tau_R \gg
\tau_C$ the second order prediction \eref{tcl2} at high temperatures
immediately becomes
\begin{equation}
\dot{p}_q(t) \approx -2 \, (1 - \cos q) \, \lambda^2 \, R_2 \,
p_q(t) \label{tcl2_diffusion}
\end{equation}
and the comparison with \eref{diffusion2} clearly shows diffusive
behavior with a diffusion constant ${\cal D} = \lambda^2 \, R_2$.
Due to the independence of $R_2$ from $n$ (again for $n > 30$) the
pertinent diffusion constant for arbitrarily large systems may be
quantitatively inferred from a finite, e.g., $30 \times 30$ layer,
see \fref{figure_D}.
Therein $\cal D$ is evaluated for the range of $\sigma$ where the
used approximation for the $q$-dependence of the correlation
function turns out to be justified, cf.~\eref{correlation1} and
\eref{correlation2}. We additionally indicate already the validity
range of the second order prediction at high temperatures, although
this point is firstly discussed in detail in the next
\sref{validity}. However, within the validity range there indeed is
a corridor of $q$ in which the dynamics at high temperatures can be
described as diffusive in terms of \eref{tcl2_diffusion}. Outside
the validity range such a $q$-corridor does not exist, since either
diffusion constants become highly energy dependent ($\sigma < 0.2$)
or localized contributions become non-negligible ($\sigma > 2$),
cf.~\fref{figure_sketch}.
\\
As indicated in \fref{figure_D}, at the l.h.s.~of the validity range
the diffusion coefficient simply scales as ${\cal D} \propto
1/\sigma^{2}$. Because such a scaling is expected in the limit of
weak disorder, we quantitatively compare with the result
$2.5/\sigma^2$ for $E = 0$ ($\lambda = 1$) according to the
Boltzmann equation, although the weak disorder limit is reached by
no means, even for the energy regime around $E = 0$. The
surprisingly good agreement supports the line of thoughts in
\sref{weak}, namely, the diffusion constants at all energies are
rather close to the value $2.5/\sigma^2$, when the disorder becomes
non-weak. At the r.h.s.~of the validity range the diffusion
coefficient begins to deviate from a simple $1/\sigma^2$-scaling,
e.g., ${\cal D} \approx 1.05 \, \lambda^2$ for $\sigma = 2$. This
value excellently agrees with the one in \cite{brndiar2008} which is
found therein by a numerical study of Green's functions. The study
in \cite{brndiar2008} remarkably requires an ensemble average over
very many realizations of disorder, while a single disorder
realization is adequate here, i.e., the correlation function is a
self-averaging object.
\\
So far, we have characterized the diffusive regime. We now turn
towards an investigation of the ballistic regime.
\begin{figure}[htb]
\centering
\includegraphics[width=0.6\linewidth]{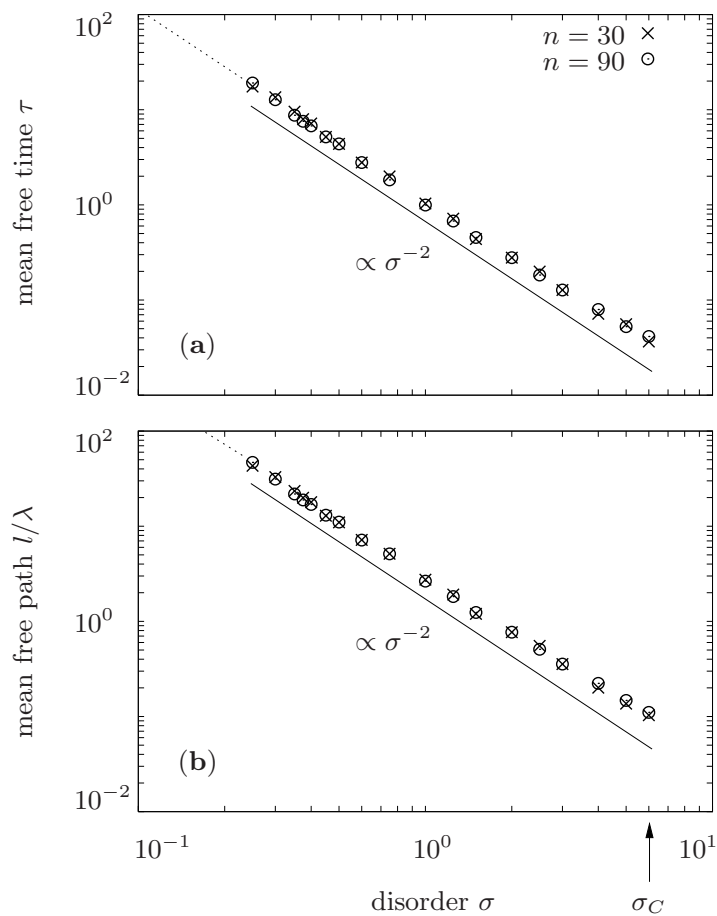}
\caption{Theoretical prediction of lowest order TCL for an
energy-independent ({\bf a}) mean free time $\tau$ and ({\bf b})
mean free path $l$ w.r.t.~the disorder $\sigma$ (Gaussian
distribution) for the layer sizes $n = 30$ (crosses) as well as $n =
90$ (circles). As a guide for the eyes proportionalities to
$\sigma^{-2}$ are shown (solid lines). The predictions for $E = 0$
according to the Boltzmann equation are also shown (dotted lines).
The results in ({\bf a}) and ({\bf b}) suggest a mean velocity
$v/\lambda \approx 2.5$.} \label{figure_l}
\end{figure}
Obviously, the replacement of \eref{tcl2} by \eref{tcl2_diffusion}
is only self-consistent for $\tau_R \gg \tau_C$, i.e., if the
relaxation time is much larger than the correlation time. But this
criterion has to break down for sufficiently large $q$, see
\fref{figure_sketch}. Hence, a transition from the diffusive to the
ballistic regime appears for those modes $p_q(t)$ which decay on an
intermediate time scale $\tau_R \approx \tau_C$, while the ballistic
regime is finally reached for those modes $p_q(t)$ which decay on a
short time scale $\tau_R \ll \tau_C$. In fact, the rate $R_2(t)$ is
found to increase linearly for $t < \tau_C/2\pi$. Due to this linear
increase the second order prediction at high temperatures is a
Gaussian decay of the corresponding modes. Such a Gaussian decay is
known to be typical for ballistic dynamics
\cite{steinigeweg2007,michel2007,steinigeweg2009}. However, the
ballistic character of the dynamics is most convincingly
demonstrated in terms of the variance $\textnormal{Var}(t)$, as
defined in \eref{variance}. The second order prediction at high
temperatures for this variance is given by
\begin{equation}
\textnormal{Var}_2(t) = 2 \, \lambda^2 \int_0^t \textnormal{d}t' \,
R_2(t')
\end{equation}
and scales as $\textnormal{Var}_2(t) \propto t^2$ for $t < \tau_C/2
\pi$. Of course, this result suggests the mean free time $\tau =
\tau_C/ 2 \pi$. Consequently, in order to obtain the mean free path
as well, a mode $p_q(t)$ with $\tau_R = \tau$ has to be considered,
i.e., the condition $p_q(\tau) / p_q(0) = 1/e$ has to be fulfilled.
This condition and \eref{tcl2} yield
\begin{equation}
2 \, (1- \cos q) \, \lambda^2 \, \int_0^\tau \textnormal{d}t' \,
R_2(t') = 1
\end{equation}
which can be rewritten as $(1- \cos q) \, \textnormal{Var}_2(\tau) =
1$. The use of $l=\pi/q$ eventually leads to the expression for the
mean free path
\begin{equation}
l = \frac{\pi}{\arccos [1 - 1/\textnormal{Var}_2(\tau)]}
\end{equation}
or the approximation $l \approx \pi/\sqrt{2} \,
\sqrt{\textnormal{Var}_2(\tau)}$ for $l \gg 1$, i.e., about two
times the standard deviation. We use this approximation, because the
dependence on $\lambda$ becomes trivial, namely, $l \propto
\lambda$. However, whenever $l \approx 1$ or even smaller, the
concrete value of the mean free path is less important, since then
the ballistic regime is restricted to a length scale below a single
lattice site and does simply not exist.
\\
As indicated in \fref{figure_l}, both the mean free time $\tau$ and
the mean free path $l$ are proportional to $1/\sigma^2$ over the
full range of accessible $\sigma$ where the used approximation for
the $q$-dependence of the correlation function turns out to be
justified, cf.~\eref{correlation1} and \eref{correlation2}.
Therefore the mean velocity $v = l/\tau$ becomes independent from
$\sigma$. Again there is a quantitative agreement with the
prediction for $E = 0$ ($\lambda = 1$) according to the Boltzmann
equation, wide outside the weak disorder limit. In contrast to
\fref{figure_D}, the validity range of the second order prediction
at high temperatures is not indicated in figure \ref{figure_l},
since this prediction for short times is expected to be valid for
all accessible $\sigma$, see the next \sref{validity}. However, for
$\sigma > 1.5$ the mean free path $l$ takes on values which are
smaller than $\lambda$, e.g., for $\lambda = 1$ the ballistic regime
is practically absent.

%
%

\subsection{Validity range of the TCL-based theory} \label{validity}
In the present section we are going to discuss the validity range of
the second order prediction at high temperatures in more detail. To
this end we consider the ratio ${\cal R}(t)$ of the fourth order to
the second order, namely,
\begin{equation}
{\cal R}(t) = \frac{\lambda^4 \, R_{4,q}(t)}{2 \, \lambda^2 \,(1 -
\cos q) \, R_{2}(t)} \, ,
\end{equation}
cf.~\eref{tcl}. Whenever ${\cal R}(t) \ll 1$, the second order term
$2 \, \lambda^2 \, (1 - \cos q) \, R_{2,q}(t)$ dominates the decay
of the modes $p_q(t)$ and the fourth order term $\lambda^4 \,
R_{4,q}(t)$ is negligible. But in general already the direct
evaluation of the fourth order term turns out be extremely
difficult, both analytically and numerically. However, by the use of
the techniques in \cite{steinigeweg2009, bartsch2008, steinigeweg2010}
the fourth order term can be approximated by
\begin{equation}
\lambda^4 \, R_{4,q}(t) \approx 4 \, \lambda^4 \, (1 - \cos q)^2 \,
R_4(t)
\end{equation}
with the remaining $q$-independent rate
\begin{equation}
R_4(t) = t \, \left [ \frac{1}{n^2} \sum_i \left (\int_0^t
\textnormal{d}\Delta \, \langle \psi_i | \, \hat{v}_0(t) \,
\hat{v}_0(t') \, | \psi_i \rangle \right)^2 - R_2(t)^2 \right ] \, ,
\label{tcl4}
\end{equation}
where $| \psi_i \rangle$ denote the eigenstates of $\hat{H}_0$.
Consequently, in complete analogy to the rate $R_2(t)$, also the
rate $R_4(t)$ may be evaluated from the consideration of an
arbitrarily chosen junction of two layers. The local interaction
between these representative layers is still called $\hat{v}_0$. The
above approximation is based on the fact that the interaction
$\hat{V}$ features the so-called Van Hove structure \cite{vanhove1954,
vanhove1957}, i.e., $\hat{V}^2$ essentially is a diagonal matrix (in
the eigenbasis of $\hat{H}_0$). However, for the concrete derivation
of this approximation we refer to \cite{steinigeweg2009} and concentrate
on the implications here. By the use of the approximation the ratio
${\cal R}(t)$ can be rewritten as
\begin{equation}
{\cal R}(t) \approx {\cal Q} \, \frac{R_4(t)}{R_2(t)} \, , \quad
{\cal Q} = 2 \, \lambda^2 \, (1 -\cos q) \, . \label{ratio}
\end{equation}
This ratio is a monotonically increasing function of $t$,
cf.~\eref{tcl4}. As a consequence there always exists a time $t_B$
with ${\cal R}(t_B) = 1$, i.e., a time where the contributions
$R_2(t)$ and $R_4(t)$ are equally large. But this fact does not
restrict the validity of the second order prediction, if $t_B \gg
\tau_R$ and hence ${\cal R}(\tau_R) \ll 1$. The validity obviously
breaks down only in the case of, say, ${\cal R}(\tau_R) \approx 1$
or even larger. Since both ${\cal R}(t)$ and $\tau_R$ depend on
$\cal Q$, we use again the condition $p_q(\tau_R)/p_q(0) = 1/e$,
i.e.,
\begin{equation}
{\cal Q} \int_0^{\tau_R} \textnormal{d}t' \, R_2(t') = 1
\label{tauR}
\end{equation}
in order to replace $\cal Q$ in \eref{ratio}. Due to this
replacement ${\cal R}(\tau_R)$ becomes a function
\begin{equation}
{\cal R}(\tau_R) = \frac{R_4(\tau_R)}{R_2(\tau_R) \, \int_0^{\tau_R}
\textnormal{d}t' \, R_2(t')}
\end{equation}
of the free variable $\tau_R$. ($\tau_R$ still depends on $\cal Q$,
of course.) Because also ${\cal R}(\tau_R)$ increases monotonically,
we define $\textnormal{max}(\tau_R)$ as the maximum $\tau_R$ for
which ${\cal R}(\tau_R)$ is still smaller than $1$. This maximum
relaxation time already specifies the validity range of the second
order prediction. However, it is useful to set
$\textnormal{max}(\tau_R)$ in relation to the correlation time
$\tau_C$.
\begin{figure}[htb]
\centering
\includegraphics[width=0.5\linewidth]{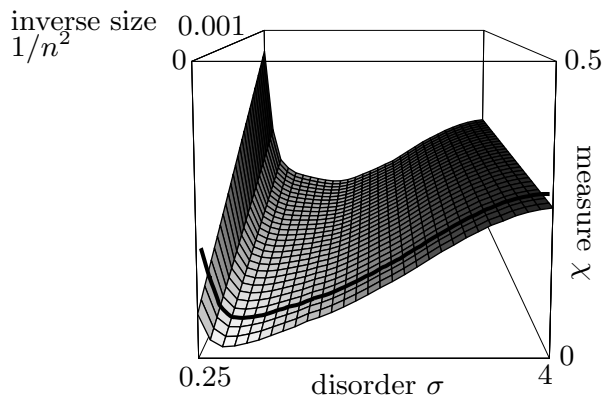}
\caption{The measure $\chi$ for the validity range of lowest
order TCL as a function of the disorder $\sigma$ and the inverse
layer size $1/n^2$. This measure detects the ``corridor'' of wave
numbers, where transport is diffusive at almost all energies with a
single diffusion constant, cf.~Fig.~\ref{figure_sketch}. For those
values of $\chi$ which are on the order of $1/4$ the corridor does
not exist. But for those values of $\chi$ which are closer to zero
the corridor opens. The smaller $\chi$, the larger is this corridor.
An absolute minimum $\chi_\textnormal{min} \approx 1/50$ is found at
$\sigma \approx 1/2$ in the limit $n \rightarrow \infty$. Note that
only $10 \%$ of the whole area is extrapolated (the area in front
of the thick line).}
\label{figure_chi}
\end{figure}
We therefore define the measure $\chi$ as the dimensionless quantity
$\chi = \tau_C/\textnormal{max}(\tau_R)$, e.g., $\chi = 1$ directly
implies the breakdown of the second order prediction on relatively
short time scales on the order of $\tau_C$, whereas $\chi = 0$
strongly indicates its unrestricted validity. For practical purposes
an interpretation of $\chi$ in the context of length scales
certainly is advantageous. Such an interpretation essentially
requires the inversion of \eref{tauR}. In general this inversion can
only be done by numerics. But we have $\tau_R = 1/({\cal Q} R_2)$
for $\tau_R \gg \tau_C$ and may hence write
\begin{equation}
\frac{1}{{\cal Q}_\textnormal{max} \,  R_2} = 2 \, \tau_C \, , \quad
\frac{1}{{\cal Q}_\textnormal{min} \, R_2} =
\frac{\textnormal{max}(\tau_R)}{2} \, ,
\end{equation}
where the factors $2$ and $1/2$ are chosen to slightly fulfill
$\tau_C \ll \tau_R \ll \textnormal{max}(\tau_R)$, i.e., ${\cal
Q}_\textnormal{max}$, ${\cal Q}_\textnormal{min}$ correspond to
lower, respectively upper border of the diffusive corridor,
cf.~\fref{figure_sketch}. We finally end up with
\begin{equation}
\frac{{\cal Q}_\textnormal{min}}{{\cal Q}_\textnormal{max}} = 4 \chi
\end{equation}
or by the use of ${\cal Q} \approx q^2 \, \lambda^2$ with
$q_\textnormal{min}/q_\textnormal{max} \approx 2 \sqrt{\chi}$. Thus,
for those values of $\chi$ which are on the order of $1/4$ the
corridor does not exist. But for those values of $\chi$ which are
closer to zero the corridor opens. The smaller $\chi$, the larger is
this corridor.
\\
In \fref{figure_chi} the measure $\chi$ is quantitatively evaluated
as a function of the amount of disorder $\sigma$ and the inverse
layer size $1/n^2$. (The rate $R_4(t)$, other than the rate
$R_2(t)$, scales significantly with $n$. This scaling gives rise to
the $n$-dependence of $\chi$.) For each layer size there is a
optimum disorder where $\chi$ is minimized, i.e., where the
diffusive corridor is maximized. But for $n = 30$ (back of
\fref{figure_chi}) we find $2 \sqrt{\chi} \approx 2/3$ at the
optimum disorder. This value indicates a corridor of about one or
two diffusive modes (for $N = 30$). For all $\sigma$ and $n \leq
100$ (which is the limit for our numerics) $\chi$ clearly appears to
be of the form
\begin{equation}
\chi(\sigma,n) = \frac{A(\sigma)}{n^2} + B(\sigma) \, .
\end{equation}
The extrapolation of the $1/n^2$-scaling eventually leads to a
suggestion for $n = \infty$ (front of \fref{figure_chi}). According
to this suggestion, we find $2 \sqrt{\chi} \approx 2/7$, again at
the optimum disorder. This value indicates a still narrow but
existent corridor of diffusive modes (for N = $\infty$).
\\
We finally recall that these findings apply at infinite temperature,
i.e., the narrow diffusive corridor is characterized by the fact
that the dynamics within this corridor is diffusive at almost all
energies with a single diffusion coefficient. The narrowness of this
corridor passes into a complete absence, since either diffusion
constants become highly energy dependent ($\sigma < 0.2$) or
localized contributions become non-negligible ($\sigma > 2$),
cf.~\fref{figure_sketch}.

%
%

\subsection{Numerical verification} \label{numerical}
In the last two sections we have introduced the TCL-based method and
have discussed its predictions as well as the validity of these
predictions. In the present section we are going to present the
results of numerical simulations in order to verify the predictions
of the method, as far as possible from the consideration of a finite
system. Since the applicability of the method requires a system
which consists of layers with a minimum size of $n = 30$, we
consider layers of that size in the following simulations. According
to the predictions of the method, for $n = 30$ a diffusive corridor
is only existent for disorders in the vicinity of $\sigma = 1$, see
\fref{figure_chi}. Therefore we focus on such a value of $\sigma$ in
all numerical simulations.
\\
For $\sigma = 1$ the TCL-based theory predicts a diffusion constant
${\cal D} \approx 2.9 \, \lambda^2$ and a mean free time $\tau
\approx 1.1$, i.e., a correlation time $\tau_C = 2 \pi \, \tau
\approx 6.9$, cf.~figures \ref{figure_D} and \ref{figure_l}.
\begin{figure}[htb]
\centering
\includegraphics[width=0.5\linewidth]{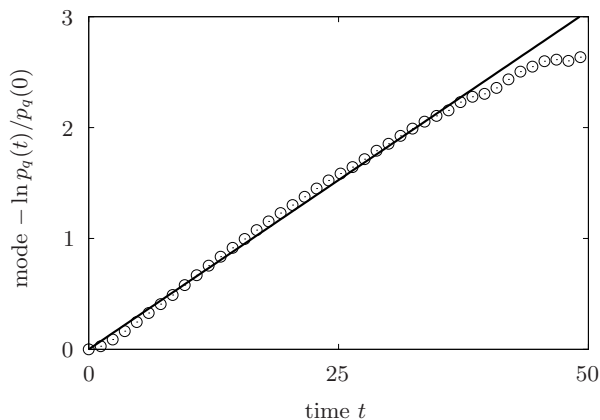}
\caption{Results for the time evolution of a mode $p_q(t)$ with $q
= \pi/20$ (without any restriction to energy regimes). The
numerical data (circles) is obtained by the use of a $4$th order
Suzuki-Trotter integrator for the model parameters $N = 40$, $n =
30$, $\sigma = 1$ (Gaussian distribution) and $\lambda = 1$
(isotropic hopping constants). The theoretical prediction of
lowest order TCL (solid line) is indicated for comparison.}
\label{figure_ST4}
\end{figure}
According to the theory, diffusive dynamics emerges only on a time
scale which is given by the condition $\tau_C \ll \tau_R = 1/[2 \,
(1 -\cos q) \, {\cal D}]$, e.g., a sufficiently small $q$ has to be
chosen. For the naturally interesting isotropic case of $\lambda =
1$ the choice $q = \pi/20$ leads to the ratio $\tau_R / \tau_C
\approx 2.0$. Unfortunately, $q = \pi/20$ is firstly realized for a
system which consists of $N = 40$ layers and such a system already
is too large for the application of numerically exact
diagonalization.
\\
However, approximative numerical integrators may be applied, e.g.,
on the basis of a Suzuki-Trotter decomposition of the time evolution
operator \cite{trotter1959,suzuki1990,steinigeweg2006-1}. In detail we choose
a pure initial state $| \psi_q(0) \rangle$  and apply a fourth order
Suzuki-Trotter integrator in order to obtain the time evolution $|
\psi_q(t) \rangle$ of this initial state and to evaluate the actual
expectation value $p_q(t) = \langle \psi_q(t) | \, \hat{p}_q \, |
\psi_q(t) \rangle$. In particular we choose the initial state at
random and only require the condition $\langle \psi_q(0) | \,
\hat{p}_{q'} \, | \psi_q(0) \rangle = \delta_{q,q'}$, i.e., we still
consider a harmonic density profile. The result of the approximative
numerical integrator is shown in \fref{figure_ST4} for a single
realization of  $| \psi_q(0) \rangle$ with $q = \pi/20$. Apparently,
there is a very good agreement between this result and the
prediction of the TCL-based theory. The latter agreement further
demonstrates that the validity of the theoretical prediction is not
restricted to an initial density matrix of the strict form $\rho(0)
= \hat{p}_q$. This fact may be understood in terms of dynamical
typicality \cite{goldstein2006,popescu2006,reimann2007,bartsch2009,
gemmer2004}.
\\
Although the above Suzuki-Trotter integrator allows to determine the
time evolution of pure initial states for rather large systems, this
integrator is not able to resolve the energy dependencies of the
dynamics, of course. To this end we have to use numerically exact
diagonalization which is applicable to a maximum system with about
$N = 10$ layers.
\begin{figure}[htb]
\centering
\includegraphics[width=0.5\linewidth]{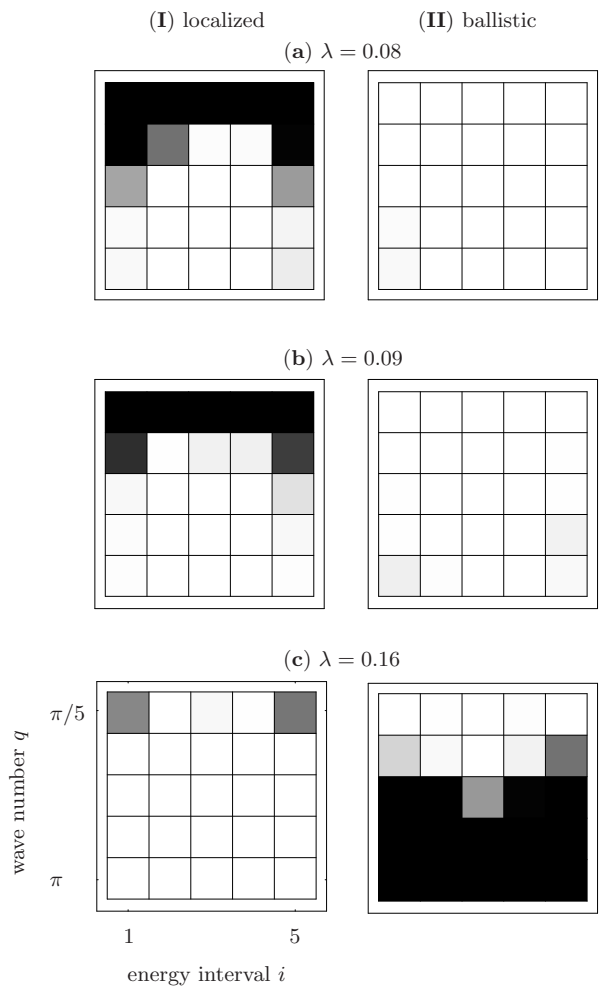}
\caption{Numerical results for the deviation of transport  from
diffusive towards ({\bf I}) localized and ({\bf II}) ballistic
behavior as a function of the wave number $q$ and the energy
interval $i$ for inter-layer hopping constants ({\bf a}) $\lambda
= 0.08$, ({\bf b}) $\lambda = 0.09$ and ({\bf c}) $\lambda =
0.16$. The color palette is chosen from white (small deviations)
to black (large deviations). The underlying data is obtained from
exact diagonalization for the model parameters $N = 10$, $n = 30$
and $\sigma = 1$ (Gaussian distribution).} \label{figure_measures}
\end{figure}
In such a system diffusive dynamics is expected to emerge only, if
the coupling constant $\lambda$ is much decreased. For $\lambda \ll
1$, in complete analogy to the above numerical simulation, the time
evolution of pure initial states has comprehensively been shown to
be in full accord with all predictions of the TCL-based
theory \cite{steinigeweg2010,steinigeweg2009}. However, since it
still remains to resolve the energy dependencies of the dynamics, we
consider the quantities
\begin{equation}
p_{q,E}(t) = \textnormal{Tr} \{ {\cal P}_E \, \rho(t) \, {\cal P}_E
\, \hat{p}_q \} \, ,
\end{equation}
where ${\cal P}_E$ denotes a projector onto the states of some
energy regime $E$. In pratice we choose a coarse-grained partition
into five energy intervals with the same number of states, namely,
there are $1800$ states in each energy interval. A fine-grained
partition into more energy intervals is not convenient, because only
a sufficiently coarse-grained partition assures $p_q(t) \approx
\sum_E p_{q,E}(t)$, i.e, the quantities $p_{q,E}(t)$ resolve the
dynamics on a reasonable energy scale. (The longer the relevant time
scale for the dynamics, the smaller is this reasonable energy scale,
of course.)
\\
The quantities $p_{q,E}(t)$ can be used in order to provide a
diagram for the dependence of transport on $q$ and $E$, similar to
the sketch in \fref{figure_sketch}. For instance, two measures for
the deviation of $ p_{q,E}(t)$ from a strictly exponential decay
(diffusive behavior) may be defined, cf.~\cite{steinigeweg2006-2}:
The first measure detects deviations towards a Gaussian decay at short
times (ballistic behavior) and the second measure detects deviations
towards a stagnant decay at long times (localized behavior). Such measures
are displayed in \fref{figure_measures}. Whenever one of these measures
is large (black areas), $p_{q,E}(t)$ does not relax exponentially.
But whenever both measures are small (white areas), $p_{q,E}(t)$
decays exponentially, i.e., it behaves diffusively. Particularly,
there indeed is a $q$-corridor where $p_{q,E}(t)$ decays
exponentially for practically all $E$. As predicted by the TCL-based
theory, this diffusive corridor is shifted to smaller $q$, when
$\lambda$ is increased. According to \fref{figure_measures} ({\bf
c}), the borders $q_{\textnormal{min}}$, $q_{\textnormal{max}}$ of
the diffusive corridor lead to a ratio
$q_{\textnormal{min}}/q_{\textnormal{max}}$ between $1/2$ and $1$.
The latter ratio remarkably is in accord with the theoretical
prediction $q_{\textnormal{min}}/q_{\textnormal{max}} \approx 2/3$,
too. (In general the ratio ${\cal Q}_{\textnormal{min}}/{\cal
Q}_{\textnormal{max}}$ has to be compared. But for the borders in
\fref{figure_measures} ({\bf c}) the approximation ${\cal Q} \approx
\lambda^2 \, q^2$ is already justified.)
\\
However, it still remains to clarify whether or not the dynamics
within the diffusive corridor is governed by a single diffusion
coefficient.
\begin{figure}[htb]
\centering
\includegraphics[width=0.5\linewidth]{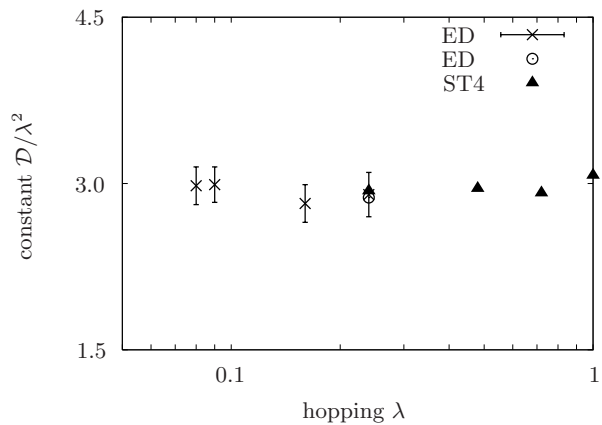}
\caption{Results for the diffusion coefficient $\cal D$ as a
function of the inter-layer hopping constant $\lambda$ for the model
parameters $n = 30$ and $\sigma = 1$ (Gaussian distribution). The
data corresponding to \fref{figure_measures} is indicated by crosses
(mean value of energy intervals) with bars (mean deviation of energy
intervals). Additional data without any restriction to energy
regimes is presented for $N = 10$ from exact diagonalization
(circle) and for $N = 10$, $20$, $30$, $40$ by the use of a $4$th
order Suzuki-Trotter integrator (triangles), cf.~\fref{figure_ST4}.}
\label{figure_Ds}
\end{figure}
To this end an exponential fit may be applied to $p_{q,E}(t)$ for
each $E$ in this $q$-corridor. Such a fit directly yields a decay
rate and consequently a diffusion coefficient ${\cal D}_E$. Then the
mean value ${\cal D}$ and the mean deviation $\delta {\cal D}$ of
the energy intervals may be evaluated, see \fref{figure_Ds}. For all
$\lambda$ where a diffusive corridor exists for $N = 10$ we find a
mean value ${\cal D}$ between $2.9 \, \lambda^2$ and $3.0 \,
\lambda^2$ and a mean deviation $\delta {\cal D}$ on the order of
less than $10 \%$. This finding finally supports the theoretical
prediction for a diffusive corridor with a single diffusion
coefficient. For completeness, \fref{figure_Ds} additionally shows
diffusion coefficients from a exponential fit to $p_q(t)$, as
obtained by the use of the above Suzuki-Trotter integrator for pure
initial states. Even though $\delta {\cal D}$ is not availabe in
that case, $\cal D$ scales simply as ${\cal D} \propto \lambda^2$
for all $\lambda$ up to the isotropic case of $\lambda = 1$.

%
%

\section{Summary and conclusion} \label{summary}

In the work at hand we have investigated single-particle transport
in the $3$-dimensional Anderson model with Gaussian on-site
disorder. Particularly, our investigation has been focused on the
dynamics on scales below the localization length. The dynamics on
those scales has been analyzed with respect to its dependence on the
amount of disorder and the energy interval. This analysis has
especially included the quantitative evaluation of the
characteristic transport quantities, e.g., the mean free path which
separates ballistic and diffusive transport regimes. For these
regimes mean velocities, respectively diffusion coefficients have
been evaluated quantitatively, too.
\\
By the use of the Boltzmann equation in the limit of weak disorder
we have shown that all transport quantities substantially depend on
the energy interval. In addition we have demonstrated that these
energy dependencies significantly differ from the well known
approximations for a free electron gas. This significant difference
develops for energies around the spectral middle where the
overwhelming majority of all states is located. As a consequence the
diffusion coefficients for these energies seem to be both a new and
relevant result.
\\
In the limit of strong disorder we have found evidence for much less
pronounced energy dependencies by an application of a method on the
basis of the TCL projection operator technique. This method suggests
that all transport quantities take on values which are practically
independent from the energy interval. Remarkably, the latter values
coincide with the prediction of the Boltzmann equation for the
spectral middle, if this prediction is simply extrapolated to strong
disorders, i.e., to disorders beyond any strict validity of the
Boltzmann equation. Solely the suggested diffusion coefficient
begins to differ from such a simple extrapolation, once the amount
of disorder becomes on the order of the critical disorder. In the
strict sense the TCL-based method does not yield a diffusion
constant in the close vicinity of the critical disorder, because the
validity range of the method is left for such an amount of disorder.
In the close vicinity of the critical disorder the diffusion
constant has to be understood as a mere conjecture. However, the
method leads to a reliable diffusion coefficient for strong
disorders which pass through almost one order of magnitude. Such a
comprehensive description appears to be novel in the literature.
\\
Strictly speaking, the TCL-based theory makes only a definite
conclusion on a corridor of finite length scales where the dynamics
is diffusive at approximately all energies with a single diffusion
coefficient. But we do not expect that the diffusion coefficient in
the diffusive regime outside this corridor is significantly
different, especially since diffusion constants should not depend on
the length scale per definition. The latter expectation is also
supported by the agreement with the Boltzmann equation and with the
numerical results for diffusion constants in \cite{markos2006, brndiar2008}.
However, whenever the above corridor of length scales
is not existent, the theory does not allow for any conclusion. Since
such a corridor may not exist in lower dimensions, the TCL-based theory
may not lead to results on transport in the one- or two-dimensional Anderson
model. But the theory itself, as demonstrated for the three-dimensional
case, can analogously be applied also to the lower-dimensional cases, of
course. This application is a scheduled project for the near future.

%
%

\ack

We sincerely thank H.-P.~Breuer and C.~Bartsch for very fruitful
discussions. Financial support by the Deutsche
Forschungsgemeinschaft is gratefully acknowledged.

%
%

\section*{References}


\begin{thebibliography}{10}

\bibitem{anderson1958}
Anderson P W 1958 {\it Phys. Rev.} {\bf 109} 1492

\bibitem{kramer1993}
Kramer B and MacKinnon A 1993 {\it Rep. Progr. Phys.} {\bf 56} 1469

\bibitem{grussbach1995}
Grussbach H and Schreiber M 1995 {\it Phys. Rev.} B {\bf 51} 663

\bibitem{slevin1999}
Slevin K and Ohtsuki T 1999 {\it Phys. Rev. Lett.}  {\bf 82} 382

\bibitem{lee1985}
Lee P A and Ramakrishnan T V 1985 {\it Rev. Mod. Phys.} {\bf 57} 287

\bibitem{abouchacra1973}
Abou-Chacra R, Thouless D J and Anderson P W 1973 {\it J. Phys.} C
{\bf 6} 1734

\bibitem{abrahams1979}
Abrahams E, Anderson P W, Licciardello D C and Ramakrishnan T V 1979
{\it Phys. Rev. Lett.} {\bf 42} 673

\bibitem{markos2006}
Marko\v{s} P 2006 {\it Preprint} arXiv:cond-mat/0609580

\bibitem{brndiar2008}
Brndiar J and Marko\v{s} P 2008 {\it Preprint} arXiv:0801.1610

\bibitem{lherbier2008}
Lherbier A, Biel B, Niquet Y-M and Roche S 2008 {\it Phys. Rev. Lett.} {\bf 100}
036803

\bibitem{dunlap1990}
Dunlap D H, Wu H-L, and Phillips P W 1990 {\it Phys. Rev. Lett.} {\bf 65} 88

\bibitem{bellani1999}
Bellani V, Diez E, Hey R, Toni L, Tarricone L, Parravicini G B, Dom\'{i}nguez-Adame F,
and G\'{o}mez-Alcal\'{a} R 1999 {\it Phys. Rev. Lett.} {\bf 82} 2159

\bibitem{peierls1965}
Peierls R E 1965 {\it Quantum Theory of Solids} (Oxford University Press)

\bibitem{kadanoff1962}
Kadanoff L P and Baym G 1962 {\it Quantum Statistical Mechanics} (Benjamin)

\bibitem{cercignani1988}
Cercignani C 1988 {\it The Boltzmann Equation and Its Applications} (Springer)

\bibitem{bartsch2010}
Bartsch C, Steinigeweg R and Gemmer J 2010 {\it  Phys. Rev.} E to be published
({\it Preprint} arXiv:1004.5364)

\bibitem{chaturvedi1979}
Chaturvedi S and Shibata F 1979 {\it Z. Phys.} B {\bf 35} 297

\bibitem{breuer2007}
Breuer H-P and Petruccione F 2007 {\it The Theory of Open Quantum Systems}
(Oxford University Press)

\bibitem{steinigeweg2007}
Steinigeweg R, Breuer H-P and Gemmer J 2007 {\it Phys. Rev. Lett.} {\bf 99} 
150601

\bibitem{michel2007}
Michel M, Steinigeweg R and Weimer H 2007 {\it Eur. Phys. J.} Special Topics
{\bf 151} 13

\bibitem{steinigeweg2009}
Steinigeweg R, Gemmer J, Breuer H-P and Schmidt H-J 2009 {\it Eur. Phys. J.} B
{\bf 69} 275

\bibitem{weaver2006}
Weaver R 2006 {\it Phys. Rev.} E {\bf 73} 036610

\bibitem{bartsch2008}
Bartsch C, Steinigeweg R and Gemmer J 2008 {\it Phys. Rev.} E {\bf 77} 011119

\bibitem{steinigeweg2010}
Steinigeweg R and Gemmer J 2010 {\it Physica} E {\bf 42} 572

\bibitem{vanhove1954}
Van Hove L 1954 {\it Physica} {\bf 21} 517

\bibitem{vanhove1957}
Van Hove L 1957 {\it Physica} {\bf 23} 441

\bibitem{trotter1959}
Trotter H F 1959 {\it Proc. Am. Math. Soc.} {\bf 10} 545

\bibitem{suzuki1990}
Suzuki M 1990 {\it Phys. Lett.} A {\bf 146} 319

\bibitem{steinigeweg2006-1}
Steinigeweg R and Schmidt H-J 2006 {\it Comp. Phys. Comm.} {\bf 174} 853

\bibitem{goldstein2006}
Goldstein S, Lebowitz J L, Tumulka R and Zanghi N 2006 {\it Phys. Rev. Lett.}
{\bf 96} 050403

\bibitem{popescu2006}
Popescu S, Short A J and Winter A 2006 {\it Nature Phys.} {\bf 2} 754

\bibitem{reimann2007}
Reimann P 2007 {\it Phys. Rev. Lett.} {\bf 99} 160404

\bibitem{bartsch2009}
Bartsch C and Gemmer J 2009 {\it Phys. Rev. Lett.} {\bf 102} 110403

\bibitem{gemmer2004}
Gemmer J, Michel M and Mahler G 2010 {\it Quantum Thermodynamics: Emergence of
Thermodynamic Behavior Within Composite Quantum Systems} (Springer)

\bibitem{steinigeweg2006-2}
Steinigeweg R, Gemmer J and Michel M. 2006 {\it Europhys. Lett.} {\bf 75} 406

\end{thebibliography}

\end{document}